\documentclass{article}
\pdfoutput=1

\usepackage{arxiv}
\usepackage{cite}           
\usepackage[utf8]{inputenc} 
\usepackage[T1]{fontenc}    
\usepackage{hyperref}       
\usepackage{url}            
\usepackage{booktabs}       
\usepackage{amsfonts}       
\usepackage{nicefrac}       
\usepackage{microtype}      
\usepackage{graphicx}       
\usepackage{amsmath}

\title{An approach to gender pay equity analysis using Bayesian hierarchical regression}

\author{
  Diana M.~Cesar\thanks{www.linkedin.com/in/DianaCesar} \\
  19 Southerland Dr.\\
  East Brunswick, NJ 08816\\
  \texttt{d.m.cesar@live.iup.edu} \\
}

\begin{document}
\maketitle
\begin{abstract}
Diversity and inclusion, or D\&I, is a topic that sparks the interest of companies, research groups, and individuals alike. Recently in the United States, renewed focus has been placed on fair and equitable pay practices, which are a key component of promoting diversity in the workplace. Despite the increased demand for reliable pay equity analysis, the challenges of conducting this type of analysis on industry data have not been adequately addressed. This paper explains a few limitations of current approaches to pay equity analysis by gender and improves on them with a Bayesian hierarchical regression model. Using global workforce data from a large US semiconductor company, Micron Technology, Inc., the paper demonstrates how the model provides a holistic view of gender pay equity across the organization, while overcoming issues more common in industry data, such as small sample size and poor gender representation. When compared to a prior analysis of Micron’s U.S. workforce, this approach decreased the amount of manual review required, enabling decision makers to finalize pay adjustments across a workforce of 31,738 people within four weeks of receiving preliminary model results. 
\end{abstract}

\keywords{Bayesian \and gender \and hierarchical linear model \and pay equity}

\section{Introduction}

Pay equity is a topic that has received much attention recently in the United States. Touching both legal compliance and social responsibility, it is important not just for companies that strive to show pay equity among their employees, but also to individuals interested in fair pay. With a rising number of U.S. companies publishing their pay equity results on the web \cite{p1, p2, p3} and certain high-profile compliance investigations capturing public interest \cite{p4} it is no wonder that so much has been said about pay equity lately. 

\quad What exactly is pay equity, and why does pay equity analysis present challenges for industry? Gender pay equity occurs when two workers of opposite genders are paid equally for the same work, when all other relevant employment factors are considered. Within the context of a workforce, one common goal of pay equity analysis is to determine which groups of workers, if any, have a wage gap, so that salary discrepancies can be appropriately addressed by the business. This can pose challenges for industry that are not commonly encountered in academic research. One such challenge is when comparison groups do not meet the sample size requirements of the modeling approach, and/or they are not sufficiently represented by both genders.

\subsection{A Simple Approach}

These issues are best illustrated through a simple example. Micron is a US semiconductor manufacturer with locations in 13 countries. At the time of this analysis, Micron’s current pay equity analysis, provided by a third-party consultant, faced limitations related to data imbalance in Micron’s comparison groups. To handle the wide range of sample sizes and gender ratios in the comparison groups, diverse types of tests had been performed on subsets of the data. These varied in complexity from separate multiple regression models (for groups with sample size >= 20 and at least 5 workers of each gender) to simple t-tests (for smaller groups) to direct salary comparisons (for groups with only 1 male and 1 female). Groups containing 1 worker had been excluded from the analysis entirely. All in all, hundreds of distinct tests had been performed on the data.

\quad The disjointed nature of the study created several inefficiencies for human reviewers when attempting to make decisions based on the results. Firstly, because the results were based on several different statistical approaches, it was not clear how groups should be prioritized for evaluation. Secondly, the simpler analyses, such as the t-tests and direct salary comparisons, failed to account for all the covariates used in the multiple regression models. This created an additional responsibility for the reviewer to gather covariate data and incorporate it into decision making for small groups. 

\quad Differing levels of reliability in the result sets, the need for case-by-case review of the simpler test results, and the exclusion of single-person groups all made the findings difficult to understand or act upon in a reasonable amount of time. Even though the scope of the study was limited to the U.S. workforce, the amount of manual labor required to interpret the results placed an unreasonable burden on the review team.

\subsection{Multiple Regression in Academic Research}

A partial solution for these issues is to combine separate tests into one multiple regression model. Multiple regression makes more efficient use of imbalanced data because, unlike a disjointed testing approach, it can pool covariate data across all comparison groups. However, this does not ensure reliable estimates for comparison groups with small sample size or poor gender representation. 

\quad Despite these limitations, multiple regression is still the most widely used approach for examining pay equity in the literature. Not much academic research has been conducted into the issues of sample size or gender representation as they pertain to pay equity analysis. This is perhaps due to differences in the objectives of academic and industry pay equity researchers. Most academic research centers around measuring the wage gap in a population at large, taking covariates and/or group effects into account. On the other hand, industry pay equity analysis seeks to measure wage gaps within specific comparison groups in the population. Consequently, regression models in the literature typically include larger grouping variables than the comparison groups used in industry (e.g., department, school, or institution), as in \cite{p5, p6, p7, p8, p9}. Fewer groups means that academic pay equity data sets are generally better balanced than a corporate pay equity data set, with a larger sample size and better gender representation in each group. 

\quad As a result, the existing literature provides few suggestions for managing small or gender-imbalanced groups. In cases where these issues did arise, some researchers opted to artificially balance the data by combining raw variables into larger groups \cite{p5} or exclude groups from the analysis plan entirely, as in \cite{p10}. However, these approaches are not ideal for industry pay equity analysis, because they do not retain the full data set with its original comparison groups. 

\subsection{Prior Hierarchical Approaches}

Hierarchical linear models (HLMs) have received more attention recently in the pay equity literature for their flexibility in modeling intricated pay structures \cite{p6, p7, p11, p12, p13, p14}. A few studies have even incorporated Bayesian statistics into an HLM by using empirical Bayes estimates \cite{p5, p8, p9}. Compared to multiple regression, HLMs make more efficient use of available data, by pooling information across groups within a hierarchical structure. This has the effect of improving the accuracy of estimates for small groups, making HLMs a natural choice for imbalanced pay equity data. However, due to the nature of academic data sets, no prior studies have demonstrated the full flexibility of HLM when working with imbalanced data. 

\quad This paper describes a project to model pay equity for Micron’s global workforce, while reducing the manual review requirements of the original analysis described above. To accomplish this, a Bayesian hierarchical linear regression model was built, utilizing Monte Carlo Markov chains through the Stan program. The paper demonstrates how the Bayesian HLM approach handled an imbalanced data set, comparing it to Micron’s current approach and to a more “standard” multiple regression model. Finally, the effectiveness of the approach is shown by summarizing how it was incorporated into a recent pay equity review cycle at Micron.

\section{Methods}

\subsection{The Data}

Data was collected from Micron’s global, non-contractor workforce as of October 5, 2017. Groups for whom a required field was not captured by Micron and workers for whom a required field was missing or inaccurate were excluded. (Excluded groups: IM Flash Technologies workers, interns, tenants, partners, workers on assignment, and fixed-term employees. Part-time workers were excluded due to inconsistency in their annualized salaries. Executives were also excluded to reduce the skew of the salary distribution.) Approximately 8\% of the original rows were removed for a final data set of 31,738 individuals, almost 30\% of whom were female. 

\quad The first step was to construct factors that could be used to categorize workers into comparison groups. At Micron, roles and responsibilities are organized in a nested hierarchy of detailed jobs within global job structures. Global job structure (GJS) is a two-digit code representing a job category and responsibility level within that category. For example, “E3” represents a technical engineer at the third level. Detailed job (“job”) is a longer description of a specific job (for example, “Data Scientist 3”). A third factor, “geo,” specifies the market value of a job in each location. For example, a “Data Scientist 3” in Taiwan would be paid differently than a “Data Scientist 3” in Boise, Idaho USA. Before entering the model, GJS and job were each crossed with geo, to create factors representing unique combinations of GJS-geo and job-geo. The factors were crossed in this way for two reasons: firstly, to allow for direct inference at the comparison group level (job-geo), and secondly, for model simplicity (this choice eliminated the need to model interaction terms explicitly). 

\quad In addition to these factors, data on performance (current ratings and an average of past ratings) and time in the job was collected for each worker. Calculation details for these covariates are omitted here for brevity. Salary was chosen as the target variable, annualized in U.S. dollars, and log-transformed to further reduce skew. 

\subsection{Model Design}

A random effect was included in the model for every GJS-geo and every job-geo to represent their intercepts. The influence of gender (binary, “female”) on salary was estimated by two sets of random effects: one set of slopes to estimate the female effect within each GJS-geo and one set of slopes to estimate the female effect within each job-geo. Taken together, these two sets of slopes would be used to derive the total wage gap for any worker in the population. The covariates (recent performance, past performance, and time in job) were entered in the model as fixed effects rather than random effects, for structural simplicity. 

\begin{figure}
  \centering
  \includegraphics[width=10cm]{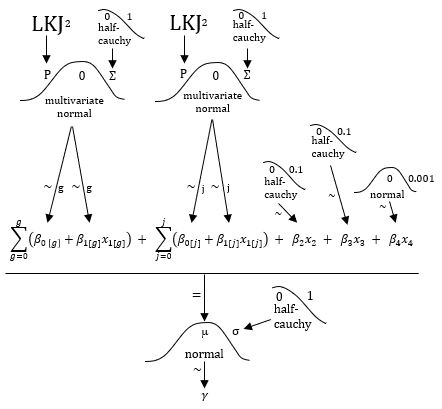}
  \caption{Hierarchical relationships between the model parameters. $\beta_0$ refers to the sets of random intercepts, one for GJS-geo and one for job-geo. $\beta_1$ represents the corresponding sets of random slopes, which estimate the effect of being female in each GJS-geo and in each job-geo. $\beta_2$, $\beta_3$, and $\beta_4$ are fixed effects for recent performance, past performance, and time in job, respectively.  Hyperparameters are represented by their mathematical symbols, depending on the distribution chosen. For random effects, square brackets [ ] next to the parameter symbol indicates the indexing scheme (either “g” for GJS-geo or “j” for job-geo). Diagram style inspired by \cite{p15}.}
  \label{fig:fig1}
\end{figure}

\quad The model was constructed with a minimally assuming linear relationship between the effects and the target variable, which is shown in the bottom half of Figure \ref{fig:fig1}. Two sets of hyperparameters with their hyperpriors were used to estimate the random intercepts and slopes corresponding to GJS-geo and job-geo. The relationship between the parameters, hyperparameters, and hyperpriors and their significance in the model are explained in the Results and Discussion. 

\subsection{MCMC in Stan}

Model training was performed in Stan, a Markov Chain Monte Carlo engine that uses a Hamiltonian Monte Carlo algorithm to construct the posterior distribution of a Bayesian model \cite{p16}. To find the model’s estimate for a given parameter, samples must be extracted from the posterior. To achieve faster convergence, priors were chosen based on the scale of parameter estimates in earlier models. For example, a normal distribution ($\mu$ = 0, $\sigma$ = 0.001), was chosen for $\beta_4$’s prior based on the tight confidence interval around the posterior mean for this parameter in an earlier model. Also, the model was specified with a non-centered parameterization, which increased the speed of the Markov chains during model training. This re-parameterization added 7,084 latent parameters to the model and brought the total number of estimated parameters up to 14,176. 

\quad Stan ran two Markov chains in parallel through 1000 warm-up and 3000 sampling iterations each, for a total run time of 63.8 hours. Convergence of the Markov chains was assessed by examining the $\hat{R}$ (Gelman-Rubin statistic), effective sample size and trace plot for the parameters estimated by the model. All $\hat{R}$s were between 1.0 (rounded to three decimal places) and 1.1, an acceptable threshold for convergence recommended by Kruschke \cite{p15}. While a handful of trace plots showed a lack of convergence within the allotted number of iterations, this was limited to less than 0.05\% of all estimated parameters. After comparing the model’s recommendations with the conclusions of the review team when presented with the same data, it was determined that this lack of convergence had no negative impact on model inference. 

\subsection{Comparison with Multiple Regression}

Because Micron’s current pay equity analysis was based on U.S. data only, a direct comparison of this approach with the Bayesian HLM was not possible. To provide a more direct comparison of results, a second, linear model (LM) was built using the same global data set as the Bayesian HLM. This model illustrates the “standard” multiple regression approach, as it commonly appears in the academic literature. It included the same factors as the Bayesian HLM minus the effects of GJS-geo. For this model, dummy variables were used in place of job-geo random effect intercepts and slope terms for the effect of being female in each job-geo were used in place of job-geo random effect slopes.

\section{Results and Discussion}

The model fit the data well, with an in-sample $R^2$ of more than 0.99 and an in-sample root mean squared error of 0.07 (both calculations based on the mean predicted log-salary for each worker). A high $R^2$ was expected, given the fact that many elements of the salary generation process were known beforehand and incorporated into the model as covariates. However, the in-sample $R^2$ of the LM was slightly higher than for the Bayesian HLM, suggesting that overfitting was a greater concern with the LM. 

\quad The model successfully estimated female effects (either zero or nonzero) for a total of 3,119 comparison groups. This was despite significant data imbalance, both in terms of group sample size and gender balance. Table \ref{tab:tab1} shows the population distribution across several factors. At the comparison group level (job-geo), more than 68\% of all groups were gender-invariant and almost 41\% of all groups consisted of a single worker. By contrast, the largest comparison group contained 826 workers. At the higher level of GJS-geo, the data imbalance was less extreme, but still significant. Almost 35\% of GJS-geos were either gender-invariant or single-worker. 

\quad In contrast, the standard multiple regression model, which had been built on the same data set, could only estimate a female effect (either zero or nonzero) for the groups that were gender variant. This amounted to 5,471 workers, over 17\% of the rows in the data set, for which manual review would have been required, because the model could not infer any female effect for their comparison group. 

\quad In addition to providing female effects estimates for all the comparison groups in the data, the Bayesian HLM also generated more robust estimates of these parameters than the LM. This improved accuracy, in comparison to the LM, was a function of the hierarchical structure of the model, shown in Figure \ref{fig:fig1}. The model is hierarchical in two respects. In the first respect, it includes random effects for two “levels” in a hierarchy: job-geo and GJS-geo. In the second respect, the model is hierarchical due to its tiered structure of parameters, hyperparameters, and hyperpriors, which helped to regularize the parameter estimates by partially pooling or shrinking them toward their overall means \cite{p17}. 

\begin{table}
 \caption{Proportion of Single-Gender and Single-Worker Groups in the Dataset}
  \centering
  \begin{tabular}{p{0.13\linewidth}p{0.24\linewidth}p{0.1\linewidth}p{0.2\linewidth}p{0.2\linewidth}}
  \toprule
    Factor     & Example     & Number of Levels     & Percent of Levels with One Gender (\%) & Percent of Levels with One Worker (\%) \\
    \midrule
    Geo         & Taiwan                    & 21        & 14.3      & 14.3    \\ 
    GJS         & E3                        & 48        & 16.7      & 6.3     \\
    GJS-Geo     & E3 Taiwan                 & 419       & 34.6      & 17.7    \\
    Job         & Data Scientist 3          & 1065      & 49.6      & 22.9    \\
    Job-Geo     & Data Scientist 3 Taiwan   & 311       & 68.2      & 40.9    \\
    \bottomrule
     
  \end{tabular}

  \label{tab:tab1}
    \begin{itemize}
 
    \item[*]\footnotesize Note: Percentages are rounded to one decimal place. Statistics for geo, GJS, and job are shown also, even though these factors were not included directly in the model, to illustrate the data imbalance of the original variables.

    \end{itemize}
\end{table}

\quad The impact of the hierarchical model structure is most noticeable in the comparison groups with small sample size. In general, the female effects estimated by the Bayesian HLM were smaller in absolute value than their corresponding LM estimates. This was particularly true for comparison groups with small sample size, where the Bayesian HLM relied more heavily on the rest of the data set to improve the estimate for the small group.

\begin{figure}[ht]
  \centering
  \includegraphics[width=12cm ,scale=0.2]{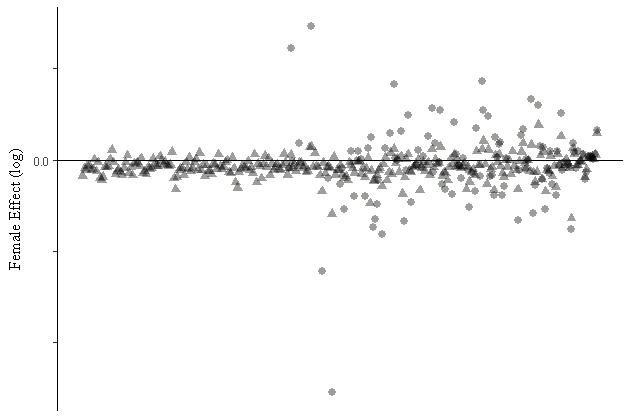}
  \caption{Mean estimated female effects from the LM (circles) and HLM (triangles), from a random sample of 300 comparison groups. Points are arranged left-to-right by group size. Values on the vertical axis are not labeled for data confidentiality.}
  \label{fig:fig2}
\end{figure}

\quad Figure \ref{fig:fig2} shows estimated female effects from a random sample of 300 comparison groups (job-geos) in the data set. For each comparison group, both the mean estimated female effect generated by the HLM and the estimated female effect from the LM (when available) are plotted. The groups are plotted in order of size from left to right. The left side of the figure is dominated by single-worker groups, for which no LM estimate is available. Starting in the middle of the plot, small groups appear. For these groups, the HLM estimates are generally much closer to the horizontal zero line (representing no gender effect) than the LM estimates. While the small sample size caused the LM to overfit the data, it had a minor impact on the estimates from the Bayesian HLM, which followed the pattern of the other group estimates in the sample. On the right-hand side of the plot where group sizes are larger, the estimates from both models show greater agreement.

Finally, wage gaps (either zero or nonzero) were estimated by generating two sets of predictions from the model for each worker: one as the worker's real gender (factual predictions) and one as the opposite gender (counterfactual predictions). From there, the predictions were converted back to the dollar scale and the difference in predictions was inspected to assess the magnitude and significance of the wage gap, if any. This procedure allowed raise recommendations to be generated for the review team to consider. It also enabled a simple calculation to measure Micron's adjusted "cents-to-the-dollar" in a manner consistent with the model results (see Equation \ref{eqn:equation1}). Adjusted cents-to-the-dollar is defined as how many cents a women earns for every dollar a man earns, when all other relevant employment factors are considered.

\begin{equation}
\label{eqn:equation1}
Adjusted \;\; Cents{\text-}to{\text-}the{\text-}Dollar =
\sum_{i=1}^{n} \hat{y}_{female} / \sum_{i=1}^{n} \hat{y}_{male}
\end{equation}
where $n$ is the number of workers in the population, $\hat{y}_{female}$ is the mean factual prediction (in dollars) when $i$ is female and the mean counterfactual prediction otherwise, and $\hat{y}_{male}$ is the mean factual prediction when $i$ is male and the mean counterfactual prediction otherwise. Since every individual is represented on both sides of the ratio, this calculation has the advantage of being balanced between genders, and thus, insensitive to the uneven gender ratios in different groups.

\section{Conclusion}

The Bayesian HLM approach greatly reduced the scope of manual review, compared to the original study. By incorporating all workers into a single model, it facilitated a top-down analysis of gender pay equity at Micron, which had previously been unfeasible given the manual review requirements of the current study. After confirming the model’s conclusions, the review team granted raises based on the model’s recommendations. From start to finish, the entire manual review activity was completed within four weeks of receiving preliminary model results and three weeks of receiving final model results.

\section*{Acknowledgement}
This project was supported by several Micron teams who offered valuable domain knowledge and technical suggestions: the HR Workforce Analytics, Enterprise Data Science, Compensation, and Employee Relations teams. 

\bibliographystyle{unsrt}  
\bibliography{references}  
\nocite{*}
\end{document}